\documentclass[%
reprint,
amsmath,amssymb,
]{revtex4-2}

\usepackage{graphicx}
\usepackage{dcolumn}
\usepackage{bm}
\usepackage{CJKutf8}
\usepackage{booktabs}
\usepackage{multirow}
\usepackage{hyperref}
\usepackage[switch]{lineno}

\begin{document}
	\preprint{APS/123-QED}
	
	\title{\textbf{The Ground State Lattice Distortion of  CsV$_{3}$Sb$_{5}$  Revealed \\by de Haas-van Alphen Oscillations}}

\author{Senyang Pan,$^{1,2}$ Feng Du,$^{3}$ Yong Zhang,$^{1}$ Zheng Chen,$^{1}$ Qi Li,$^{1,2}$ Yingcai Qian,$^{1,2}$ Xiangde Zhu,$^{1}$ Chuanying Xi,$^{1}$ Li Pi$^{1,2}$}
\author{Wing Chi Yu$^{4}$}
\email{wingcyu@cityu.edu.hk}
\author{Qun Niu$^{1}$}
\email{qniu@hfml.ac.cn}
\author{Jinglei Zhang$^{1}$}
\email{zhangjinglei@hfml.ac.cn}
\author{Mingliang Tian$^{1,5}$}

\affiliation{$^{1}$Anhui Key Laboratory of Condensed Matter Physics at Extreme Conditions, High Magnetic Field Laboratory, HFIPS, Chinese Academy of Sciences, Hefei 230031, China
\\$^{2}$University of Science and Technology of China, Hefei 230026, China\\$^{3}$Max Planck Institute for Chemistry, Hahn Meitner Weg 1, 55128 Mainz, Germany\\$^{4}$Department of Physics, City University of Hong Kong, Kowloon, Hong Kong, China\\$^{5}$Department of Physics, School of Physics and Materials Science, Anhui University, Hefei, Anhui 230601, China}%

\date{\today}

\begin{abstract}
	The recently discovered AV$_{3}$Sb$_{5}$ (A = K, Rb, Cs) compounds have garnered intense attention in the scientific community due to their unique characteristics as kagome superconductors coexisting with a charge density wave (CDW) order. To comprehend the electronic properties of this system, it is essential to understand the lattice distortions associated with the CDW order and the ground state electronic behavior. Here, we comprehensively examine the Fermi surface by a combination of angle-dependent torque magnetometry and density functional theory calculations. We observe magnetic breakdown in the de Haas-van Alphen oscillations under high magnetic fields. Additionally, by examining the angular and temperature variations in quantum oscillation frequencies, we gain insight into the evolution of the three-dimensional like Fermi surfaces and the cyclotron masses of the orbits, which are consistent with weak electron-phonon coupling. Notably, further comparisons indicate that the 2$\times$2$\times$2 Star-of-David (SoD) distortion is more compatible with both high frequency data above 1000\,T and low frequency data below 500\,T, while the 2$\times$2$\times$2 Tri-Hexagonal (TrH) distortion aligns well with experimental data at mid frequencies. This finding implies the inherent coexistence of both TrH and SoD 2$\times$2$\times$2 patterns within the CDW order. These observations provide key insights into the interplay among effective electronic dimensionality, CDW state, and superconductivity.
\end{abstract}

\maketitle

\section{\label{sec:level1}Introduction}

Metals featuring kagome lattices have gathered substantial research attention due to their unique properties, including symmetrically protected Dirac and Weyl points, van Hove singularities, and flat bands resulting from geometric features within their electronic band structures~\cite{1,2,3}. The distinctive geometry and ground states of kagome lattices make them a promising platform for investigating a wide range of quantum phenomena, including quantum spin liquids, superconductivity, charge density wave (CDW), as well as Coulomb interactions among adjacent lattice-site electrons~\cite{4,5,6,7}. \\

The recently identified kagome materials AV$_{3}$Sb$_{5}$ (A = K, Rb, Cs) exhibit a diverse array of intriguing physical phenomena closely linked to its unique lattice structure and electron behavior at the Fermi level~\cite{8,9,10,11,12,13,14,15,16,17,18,19,20,21,22,23}. As illustrated in the inset of Fig.\,1, the kagome layer is constituted of V atoms, with adjacent Sb atoms forming a triangular lattice at the hexagonal center. Another triangular lattice formed by alkali atoms are positioned above or below the layer~\cite{8,22}. In AV$_{3}$Sb$_{5}$, time-reversal symmetry breaking within the CDW phase imply its unconventional nature. Additionally, various modulations, linked to nematic fluctuations, have been identified in the CDW phase, underscoring the complex interplay of structural and electronic properties. Consequently, two in-plane superlattice modulation modes have been proposed in CsV$_{3}$Sb$_{5}$: the Star of David (SoD) and Tri-Hexagonal (TrH) modes~\cite{7,8,9,20,21,24,25,26}. Given that these novel physical properties are intimately linked to the V kagome layer, it is crucial to determine the significant impact of this lattice distortion pattern on Fermi surface (FS) and topological properties. Theoretical calculations indicate that the crystal structural reconstructions are associated with phonon softening at M and L points, giving rise to various candidate distortion modes~\cite{12,21,23,27}.  However, the experimental validations of these theories remain a subject of considerable debate. Nuclear magnetic resonance (NMR) and X-ray diffraction experiments have revealed a three-dimensional (3D) charge modulation with a 2$\times$2 period in CsV$_{3}$Sb$_{5}$~\cite{7,11,18,20}. In addition to the in-plane charge order, an extra 4\textit{c} unidirectional charge order has been observed on the surface by single crystal diffraction and certain scanning tunnelling microscopy (STM) experiments in CsV$_{3}$Sb$_{5}$~\cite{20,28}. Furthermore, STM experiments indicate an in plane 1×4 charge modulation occurring at temperatures below 50-60\,K~\cite{29,30}.\\

Quantum oscillations measurement is a powerful tool to directly probe the novel electronic states, which had found wide application in the study of superconductors and topological materials~\cite{14,15,31,32}. In this work, we present a detailed torque magnetometry study in combination with density functional theory (DFT) calculations to elucidate the ground electronic structures of CsV$_{3}$Sb$_{5}$. By plotting de Haas–van Alphen (dHvA) oscillations in magnetic fields up to 33\,T, we have identified seven distinct frequencies ranging from 19 to 2021\,T when the magnetic field is parallel to the crystallographic \textit{c} axis (\textit{B}$\parallel$\textit{c}). Notably, the high field data reveal the presence of two supplementary orbits, corresponding to the phenomenon of magnetic breakdown, i.e., tunneling of quasiparticles between adjacent pockets of the FS above a threshold magnetic field. Therefore, the distortion of FS cannot be adequately explained by theoretical calculations when only the number of frequencies is considered. We observe dHvA oscillations originating from the $\beta$, $\eta$, $\mu$ and their magnetic breakdown orbits within the basal plane. The corresponding effective masses range from 0.235 to 0.779 times the free electron mass ($m_e$). The field-angle dependence of these frequencies diverge from the expected \textit{F}$\propto$1/$\cos$$\theta$ relationship, revealing a 3D nature of FSs. Furthermore, we examined the correlation between dHvA oscillation data and theoretical calculations based on various periodic distortions. Our results indicate that the 2$\times$2$\times$2 TrH and 2$\times$2$\times$2 SoD periodic lattice distortion exhibit their own advantageous frequency ranges in interpreting angle-dependent oscillation data. These findings provide valuable insights into the ground state lattice distortion in CsV$_{3}$Sb$_{5}$, which is crucial for advancing our understanding of the family of layered kagome materials.

\section{Experiment and calculation}
High-quality CsV$_{3}$Sb$_{5}$ single crystals in our studies were synthesized by the self-flux method, and detailed information regarding the crystal growth procedure can be found in Ref~\cite{8,13}. The dHvA measurements were carried out in both the superconducting magnet (Oxford Instrument Inc.) and the DC-resistive magnet at the Chinese High Magnetic Field Laboratory (CHMFL) in Hefei. The magnetic torque was tracked by measuring the capacitance between the cantilever and a gold film, using an AC bridge (Andeen Hagerling, 2700A). The sample was fixed on a 25\,$\mu$m CuBe platform that could be rotated in situ around \textit{c} axis, as shown in the inset of Fig.2(a). The $\theta$ is defined as the angle between the magnetic field direction and the crystallographic c-axis within the \textit{ac} plane.\\

The FS calculation was performed using DFT with all-electron full-potential linearized augmented plane-wave implemented in the WIEN2k package~\cite{33}. Experimental lattice parameters for the 2$\times$2$\times$1 TrH~\cite{21}, 2$\times$2$\times$2 SoD, 2$\times$2$\times$2 TrH~\cite{34}, and 2$\times$2$\times$4 TrH~\cite{20} structures were used. In all the calculations, generalized gradient approximation of Perdew, Burke, and Ernzerhof~\cite{35} was employed as the exchange-correlation potential, and $R_{MT}^{\min}K_{\max}$ was set to 7.5. The muffin-tin radius for the Cs, V, and Sb atoms was 2.5, 2.47, 2.5 respectively. A \textit{k}-point mesh of 40000, 20000 and 10000 in the first Brillouin zone were used for the 2$\times$2$\times$1, 2$\times$2$\times$2 and 2$\times$2$\times$4 structures. The quantum oscillations frequiencies and extremal FS cross sections were then extracted by the Supercell \textit{k}-space External Area Finder code~\cite{36}.

\section{Results and discussion}

\begin{figure*}
\includegraphics[width=6in]{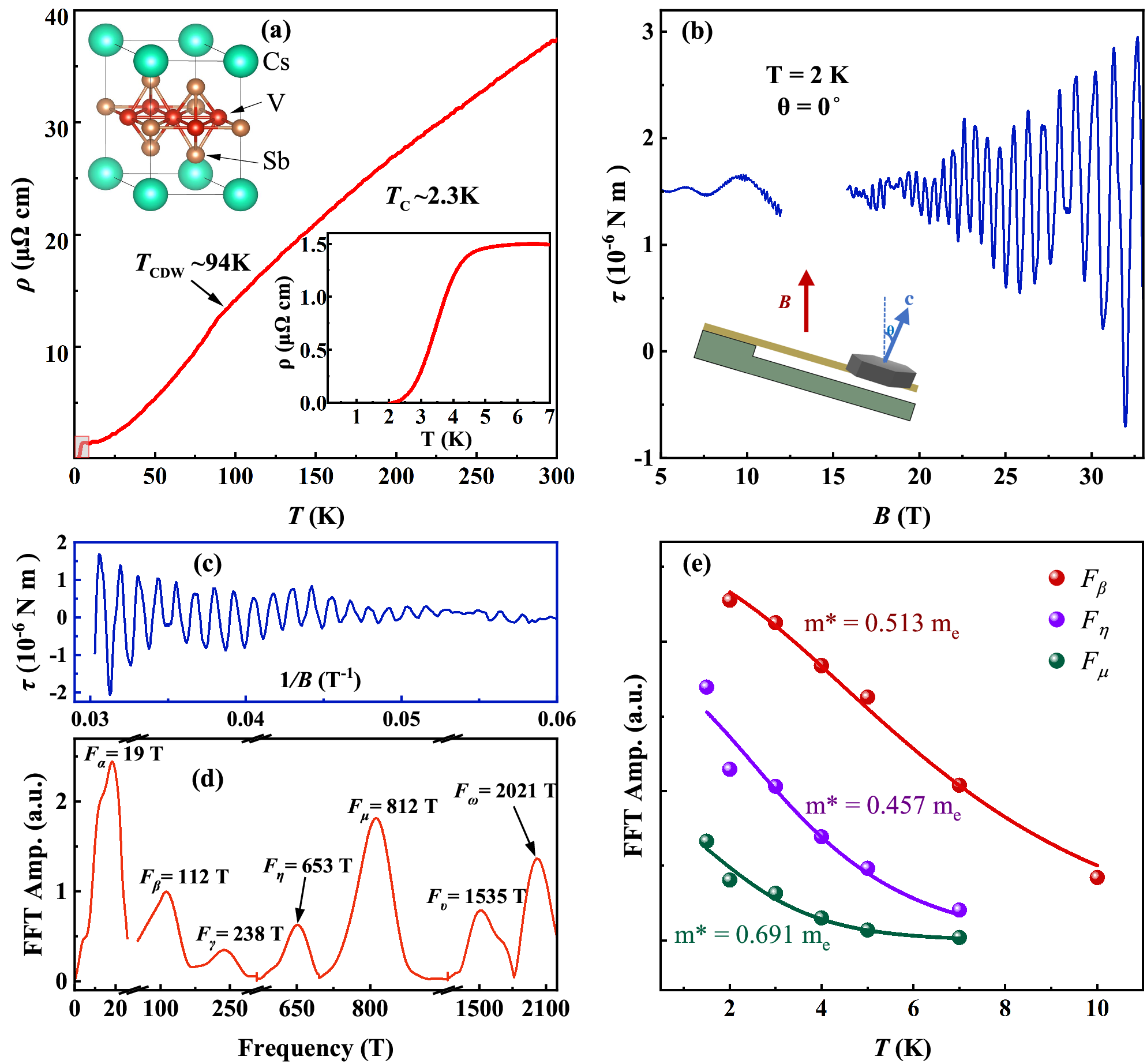}
\caption{\label{fig:wide}(Color online) (a) Temperature-dependent resistivity $\rho$ measured with the current applied along the \textit{ab} plane. (Inset) The crystal structure of ${\mathrm{CsV_{3}Sb_{5}}}$ (space group $P6/mmm$ No.191). The green, golden and red spheres represent Cs, Sb and V atoms, respectively. (b) Magnetic torque was measured at 2\,K in fields up to 33\,T, with the non-oscillatory background subtracted. (Inset) Schematic of the torque measurement setup, indicating $\theta$ as the angle between the applied field and the \textit{c} axis. (c) Oscillatory component of the torque signal as a function of the inverse magnetic field 1/\textit{B}. (d)The oscillation signal spectrum, obtained within a 20 to 33\,T field range, reveals seven distinct dHvA frequencies after undergoing FFT. (e) Temperature dependence of FFT
amplitudes of \emph{F$_\beta$}, \emph{F$_\eta$}, and \emph{F$_\mu$} peaks. The solid lines represent the LK formula fits for effective masses.}
\end{figure*}

Figure\,1(a) displays resistivity as a function of temperature, with the electrical current flowing in the crystallographic \textit{ab} plane in zero magnetic field. It exhibits a typical metallic behavior with a kink at \(T_{\mathrm{CDW}}\) \(\sim 94 \, \text{K}\), which can be attributable to the CDW transition driven by Peierls instability~\cite{18}. At a lower temperature, a sharp superconducting transition occurs at \(T_{\mathrm{c}}\) \(\sim 2.3 \, \text{K}\). Both the CDW transition temperature and $T_{\mathrm{c}}$ are consistent with the previous studies~\cite{22,35,67,38,39}. After the background is subtracted via polynomial fits, the typical quantum oscillations of CsV$_{3}$Sb$_{5}$ at 2\,K are presented in Fig$_{.}$\,1(b). Figure\,1(c) represents the dHvA oscillations in the magnetization as a function of the inverse magnetic field 1/\textit{B}. By applying a fast Fourier transform (FFT), we identify seven distinctly fundamental frequencies, which are labeled as \emph{F$_\alpha$} =19\,T, \emph{F$_\beta$}  =112\,T, \emph{F$_\gamma$}=238\,T, \emph{$F_\eta$}=653\,T, \emph{$F_\mu$}=812\,T, \emph{$F_\upsilon$}=1535\,T and \emph{$F_\omega$}=2021\,T in Fig$_{.}$\,1(d). These frequencies suggest the influence of multiple bands dominating the oscillations, and their connection to extremal cross sections of the relevant Fermi pockets via the Onsager relation ~\cite{40}.  Our findings provide a more comprehensive range of observable fundamental frequencies than those in Shubnikov–de Haas (SdH) quantum oscillation studies, aligning with recent researches ~\cite{14,24,41,42}. We notice that the new frequencies associated with large extremal orbits ranging from 2085\,T to 2717\,T have been observed in the thin flakes, which implies that the reduced dimensionality could enlarge the FSs in CsV$_{3}$Sb$_{5}$~\cite{43}.\\

Turning to the mid-frequency spectrum, previous results showed two distinct frequencies between 500\,T and 1000\,T measured by SdH quantum oscillation studies~\cite{14,42}. Our measurements in superconducting magnets exhibit a superior signal-to-noise ratio with 5 to 12\,T FFT fit window, yielding four distinct frequencies. These frequencies, labeled as \emph{$F_\eta^{\prime}$}=558\,T, \emph{$F_\eta^{\prime\prime}$}=637\,T, \emph{$F_\mu^{\prime}$}=720\,T and \emph{F$_\mu^{\prime\prime}$}=811\,T, are easily distinguishable. Furthermore, in high magnetic field these four peaks can evolve into clear two peaks of \emph{$F_\eta$} and \emph{$F_\mu$}  , as shown in the inset of Fig$_{.}$\,1(d). The values of \emph{$F_\eta$} and \emph{$F_\mu$} are closely resemble those of \emph{$F_\eta^{\prime\prime}$} and \emph{F$_\mu^{\prime\prime}$}. Additionally, they approximate the summed values of \emph{F$_\beta$} with \emph{$F_\eta^{\prime}$} and \emph{F$_\beta$} with \emph{$F_\mu^{\prime}$}, respectively. Theoretically, such an extra orbit could be observed with adjacent Fermi pockets due to quasiparticle tunneling~\cite{44}. This includes both the combinational and forbidden frequencies, arising from the magnetic breakdown of the quasi-classical approximation~\cite{38}. As illustrated schematically in Fig$_{.}$\,2(b), when magnetic field exceeds the threshold value, electrons orbiting around the FS can tunnel among different FS giving rise to larger orbit. Our dHvA results depict an intermediate case where incomplete features of magnetic breakdown are observed at low magnetic fields, coupled with novel effects induced by ``mixing''. The results in a complex frequency spectrum displaying a mix of combinational and forbidden frequencies in the low magnetic field regime, while the frequencies of quantum oscillations decreases at high magnetic fields because tunneling has fully occurred.\\

\begin{figure*}
\includegraphics[width=5.5in]{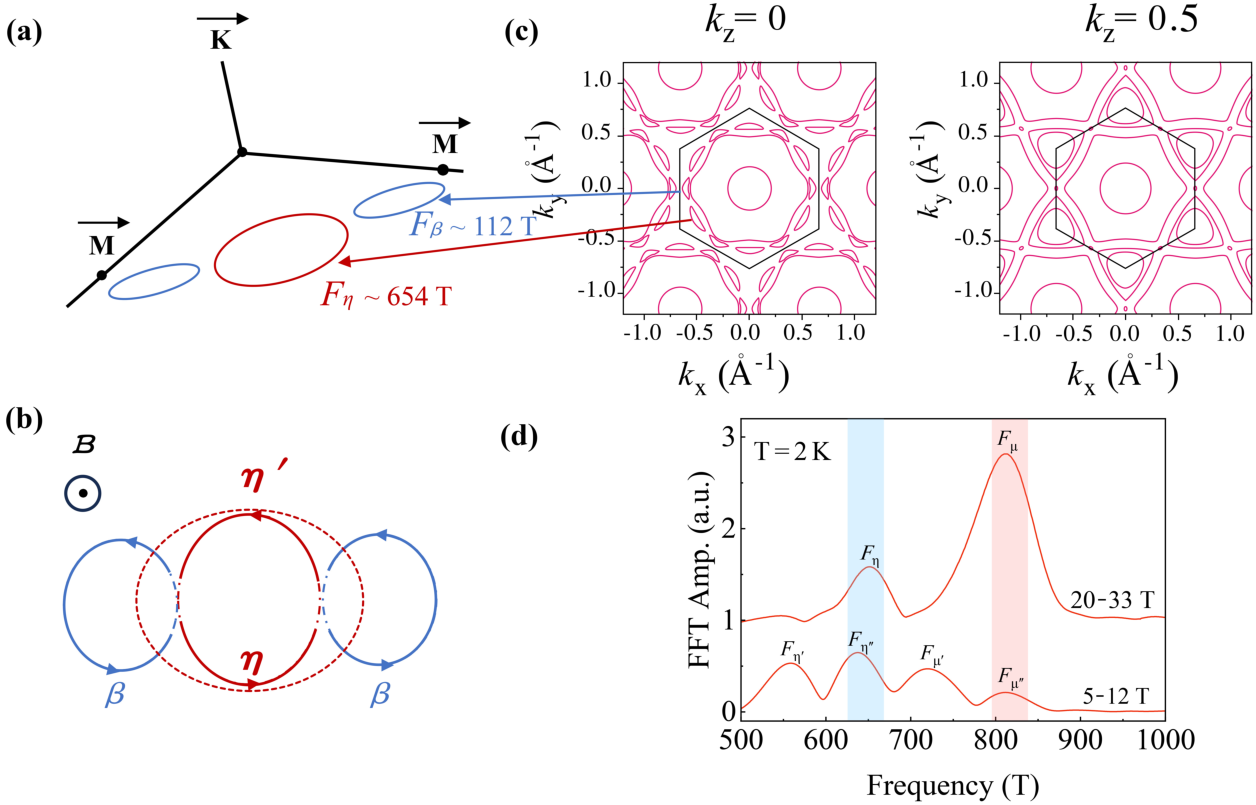}
\caption{\label{fig:wide}(Color online) (a) Projections of the $\beta$ and $\eta$ FS pockets inside the Brillouin zone. (b) The pockets are close to each other in the momentum space allowing tunneling under high field. (c) We show isoenergy contours at $k_z=0$ and $k_z=0.5$ for the $E_{\mathrm{F}}=E_{F, \exp }$ surface with unique extremal orbits marked. (d) FFT spectrum of the dHvA signal for various magnetic field range. In the low field, four peaks are observed in the mid-frequency spectrum (refer to the discussion below).}
\end{figure*}

Following the theory of magnetic breakdown, the breakdown field is proportional to the square of the gap between adjacent pockets of the FSs in \textit{k} space~\cite{21,23,44}. In general, the magnetic fields required for this particular effect are too large to be observed experimentally. However, magnetic breakdown has been detected within different frequency ranges in CsV$_{3}$Sb$_{5}$~\cite{33,46,47}; this might be attributed to the small band gap separating adjacent sections of the Fermi surface. An analogical case is observed in the Nernst oscillations with frequencies below 100\,T~\cite{34,48}, where the additional frequency is also attributed to the magnetic breakdown across the orbits. Recent data derived from pulsed magnetic fields reveals additional frequencies, which are interpreted as arising from magnetic breakdown orbits encompassing multiple triangular sheets~\cite{39}. In Fig$_{.}$\,2(c), we present the calculated FS sections based on the TrH mode at $k_z=0$ and $k_z=0.5$ respectively, where the $\eta$, $\mu$ and $\beta$ orbits are adjacent within the Brillouin zone. Among these, the $\beta$ band lies in close proximity to the $\eta$ and $\mu$ bands, which provides the possibility of magnetic breakdown. However, tunnel-diode oscillator measurements get the consistent numbers of frequencies between 500 and 1000\,T under magnetic fields up to 35\,T and 41.5\,T~\cite{24}. The additional high frequency components in their works are attributed to more complex structural distortions, instead of resulting from magnetic breakdown. Nonetheless, the magnetic breakdown effect hinders the accurate explanation of intrinsic frequency in CsV$_{3}$Sb$_{5}$ by quantum oscillations when only the number of frequencies is considered.\\

\begin{table}[ht]
\caption{\label{tab:table4}
 Comparison between calculated and measured dHvA frequencies and effective masses for different bands.}
 \begin{ruledtabular}
	\centering
	\begin{tabular}{ccccccc}
		\multirow{2}{*}{\textbf{Band}}&\multicolumn{2}{c}{\textbf{\underline{Measured}}}&\multicolumn{2}{c}{\textbf{\underline{Calculated}}}&\multirow{2}{*}{$\delta$ = ($\frac{m}{m_{c}}$ -1)$\times$100$\%$}\\
		&F(T)&$m$/$m_e$&F(T)&$m_c$/$m_e$\\
		\midrule
		$\alpha$ &19&0.199&67.5&0.146&36.3$\%$\\
		$\beta$ &112&0.235&\rule{0.8cm}{0.05em}&\rule{0.8cm}{0.05em}&\rule{0.8cm}{0.05em}\\
		$\eta$ &654&0.560&528.5&0.292&91.2$\%$\\
		$\eta^{\prime}$ &558&0.457&\rule{0.8cm}{0.05em}&\rule{0.8cm}{0.05em}&\rule{0.8cm}{0.05em}\\
		$\eta^{\prime\prime}$ &637&0.513&\rule{0.8cm}{0.05em}&\rule{0.8cm}{0.05em}&\rule{0.8cm}{0.05em} \\
		$\mu$ &812&0.779&716.7&0.417&86.8$\%$\\
	    $\mu^{\prime}$ &719&0.693&\rule{0.8cm}{0.05em}&\rule{0.8cm}{0.05em}&\rule{0.8cm}{0.05em} \\
		$\mu^{\prime\prime}$ &811&0.691&\rule{0.8cm}{0.05em}&\rule{0.8cm}{0.05em}&\rule{0.8cm}{0.05em} \\
		$\omega $ &2021&0.835&1928.8&0.842&\rule{0.8cm}{0.05em} \\
	\end{tabular}
\end{ruledtabular}
\end{table}

\begin{figure*}
\includegraphics[width=6.1in]{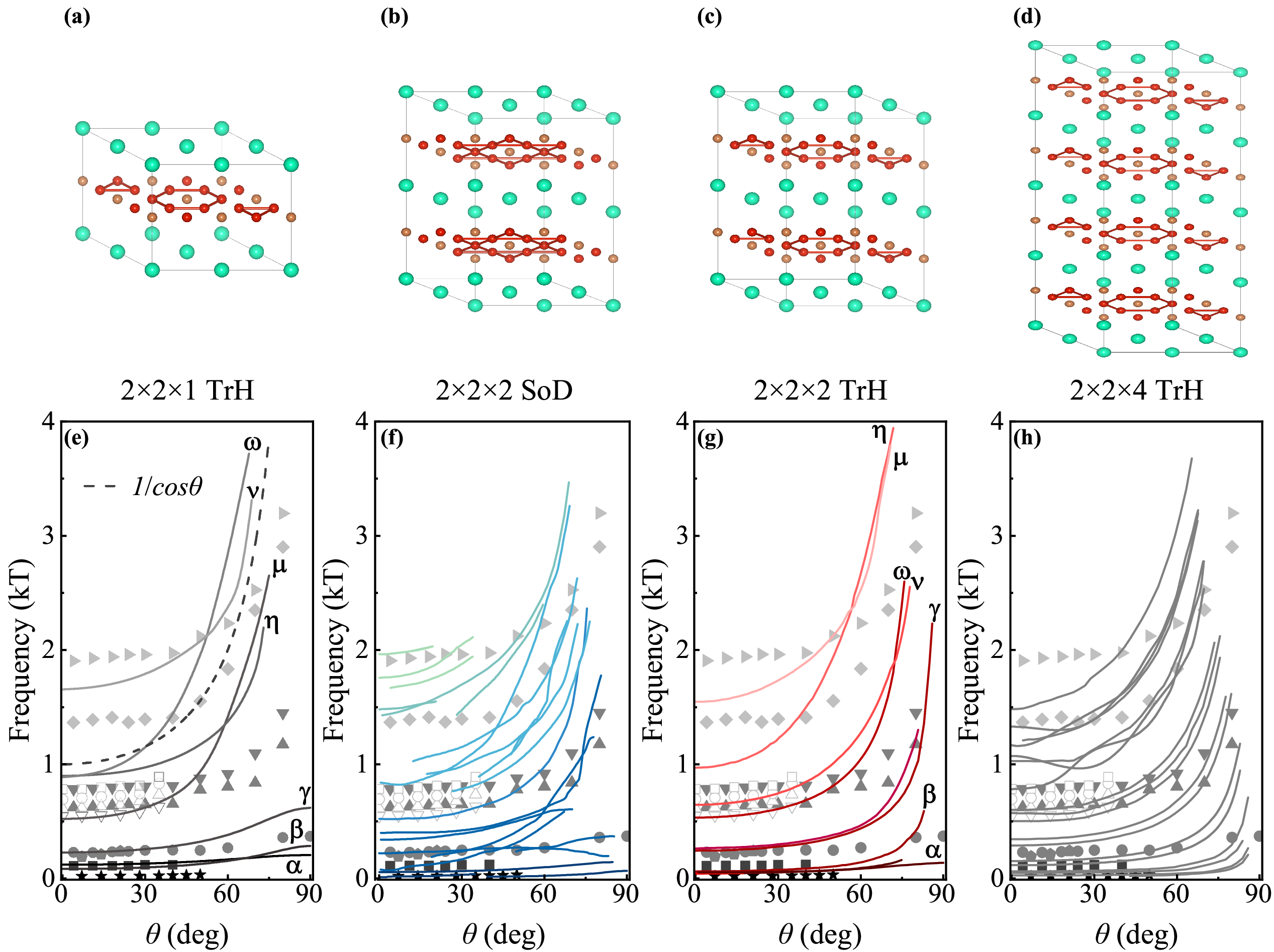}
\caption{\label{fig:wide}(Color online) (a)-(d) We observed the angular dependence of oscillation frequencies in four distorted mode supercells of CsV$_{3}$Sb$_{5}$. The images above show the corresponding lattice distortion patterns respectively. This enabled direct comparison between the experimental dHvA data (represented by solid or hollow symbols) and the DFT models. The hollow symbol represents the frequency when the low field magnetic breakdown did not occur. Frequencies represented by different shapes symbols have been grouped by the pocket of the origin.}
\end{figure*}

Basically, the magnetic torque $\tau$ is proportional to magnetization $\tau$ $\propto$ \textit{M}$\times$\textit{B}, the dHvA oscillation can be described by the Lifshitz-Kosevich (LK) formula as follows, with the Berry phase taken into account:
\begin{equation}
	\Delta \tau \propto-B^{p} R_{T} R_{D} R_{S} \sin \left[2 \pi\left(\frac{F}{B}+\gamma-\delta\right)\right]
\end{equation}

where $R_{T}=\alpha Tm^{*} /[Bm_{0}  \sinh( \alpha Tm^{*} /B m_{0})]$, $R_{D} =\exp(-14.69 m^{*}   T_{D}  / B)$, $R_{S} = \cos (\pi  g^{*}   / 2  m_{0} )$, $T_D$  is the Dingle temperature and $m^*$ being the effective mass. The $g^{*}$ is the effective g factor, $\gamma \mathit{-} \delta$  is the phase factor, in which $\gamma  =  \frac{1}{2} \mathit{-} \frac{\Phi_B}{2\pi}$ and $ \Phi_B $ is Berry phase. The $p$ and $\delta$ are determined by the dimensionality of FS: $p =  \frac{1}{2}$ and $\delta =  \pm \frac{1}{8}$ for 3D FS, $p$= 0 and $\delta$=0 for two-dimensional (2D) FS case, respectively. The inverse magnetic field 1/\textit{B} is defined as 1/\textit{B} = (1/$B_{\mathrm{max}}$  +1/$B_{\mathrm{min}}$)/2, where $B_{\mathrm{max}}$  and $B_{\mathrm{min}}$  define the range of magnetic fields used by FFT. The effective masses fits for relative oscillatory component, derived from their FFT amplitudes as a function of temperature. Table$_.$\,1 displays the effective masses corresponding to $\alpha$, $\beta$, $\eta$, $\mu$ and $\omega$ orbits are $m^*$= 0.19(9), 0.23(5), 0.56(0), 0.44(9), 0.83(5)$m_e$, respectively. For frequencies below 1000\,T, these light effective masses are close to the previous reports, while the Fermi pocket corresponding to $F_\omega$= 2021\,T has a larger value 0.83(5)$m^*$~\cite{14,20,42,47}. For the mid-frequency orbits, within experimental uncertainty, we find that $m_\eta$ and $m_\mu$ indeed agree with corresponding cyclotron masses from their individual pockets involved in breakdown orbit. The frequency in the FFT and the extracted mass are all consistent with theoretical predictions for Klein tunneling ~\cite{38,42,45}.\\

The effective mass $m^*$ for these orbits extracted from 2$\times$2$\times$2 TrH structure band calculation are compared with the measured values. Since the exclusion of the electron-phonon (el-ph) interaction in full-potential local-orbital calculations, the mass enhancement percentage~\cite{50,51} $\delta=\left(m / m_c-1\right) \times 100\,\%$ can be extracted from experimental, m, and theoretical, $m_c$, effective masses. For the $\alpha$, $\eta$ and $\mu$ bands, $\delta$ obtained from both theoretical calculations and measurements are almost the same. As for the $\omega$ band, the experimentally measured effective mass value is almost identical to those theoretical calculation. These minimal el-ph effects support the notion of conventional superconducting in CsV$_{3}$Sb$_{5}$, which is consistent with recent Sb$^{121}$ nuclear quadrupole resonance (NQR) spectra and magnetic penetration depth measurements~\cite{13,25,52}.\\ 

To clarify the morphology of the FSs in CsV$_{3}$Sb$_{5}$, we performed angle-dependent quantum oscillation measurements with the magnetic field rotated from the \textit{B}$\parallel$\textit{c} axis to the \textit{ab} plane. We analyzed the frequencies of the peaks obtained from the FFT spectrum against the tilted angle. The frequencies, denoted by red hollow symbols, are obtained from the lower field before the magnetic breakdown occurs. In Fig$_{.}$\,3, we noticed that the quantum oscillations observed at higher tilted angle start to deviate from 1/$\cos$\,$\theta$ kT (guided by the gray dotted line), implying a 3D character in the FS pockets.\\

In Fig$_{.}$\,3(e)-(h), we conducted a comparative analysis between the experimental data and the calculated oscillatory frequencies derived from the cross-sectional areas of FSs based on four distortion modes as illustrated in Fig$_{.}$\,3(a)-(d). These modes encompass 2$\times$2$\times$1 TrH, 2$\times$2$\times$2 TrH, 2$\times$2$\times$2 SoD, and 2$\times$2$\times$4 TrH supercell structures, each characterized by unique lattice distortions. Our analysis predominantly focuses on the layer of V and Sb atoms that constitute the kagome plane, while ignoring the associated Sb atomic layer. It is clear that the nature of their out-of-plane lattice distortions involve subtle interactions and changes in the FS morphology. Notably, the 2$\times$2 mode demonstrates a superior fit with the experimental data, particularly in the context of frequency values and their evolution with changes in the magnetic field tilt angle. As depicted in Fig$_{.}$\,3(f), the 2$\times$2 SoD theoretical computations align well with the observed data, while resulting in additional frequencies. It implies that the more frequencies increases the probability of agreement with experimental data, and the frequency evolution with angles should be considered. Moreover, quantum oscillation frequencies are well fitted until the angle increased to 70$^{\circ}$ for the 2$\times$2 TrH mode in Fig$_{.}$\,3(g). The 2$\times$1 and 2$\times$4 TrH modulation mode curves fail to accurately represent the ground state lattice distortion in CsV$_{3}$Sb$_{5}$.\\

For the $F_\eta$ and $F_\mu$ pockets, 2$\times$2$\times$2 TrH modulation exhibits a stronger correspondence with the experimental data. However, the two highest frequencies observed in the experiment can be basically fitted by the modulation of 2$\times$2$\times$2 SoD lattice modulation. At the same time, it is difficult to distinguish between 2$\times$2$\times$2 TrH and 2$\times$2$\times$2 SoD lattice modulation in experimental fitting for frequencies below 500\,T. The results indicate that structural distortions have less influence on the smaller Fermi pockets. Our observations are in favor of a coexistence of the 2$\times$2$\times$2 TrH and 2$\times$2$\times$2 SoD lattice modulation in CsV$_{3}$Sb$_{5}$ at the base temperature, which is evidently associated with the morphology of the FSs.\\

Based on our dHvA results, we argue that the 3D-like FSs could be attributed to the interlayer ordering along the \textit{c} axis direction.  Recent quantum oscillations data have reported quasi-2D FSs, especially for high-frequency peaks over 2000\,T~\cite{14,42,43}. Both STM measurements and X-ray diffraction data have demonstrated 2$\times$2$\times$2 and 2$\times$2$\times$4 modulation, revealing both TrH and SoD like distortions~\cite{28,29,48,52}. Moreover, studies on the control of microscopic materials show that the FS of CsV$_{3}$Sb$_{5}$ is sensitive to magnetic field, stress, and even differences between samples~\cite{53,54}.  The 3D nature of FS becomes evident in scenarios focusing on out-of-plane reconstruction with a prevalent 2$\times$2$\times$2 pattern dominance. In contrast, 2$\times$2$\times$4 systems with out-of-plane reconstruction might exhibit prominent quasi-two-dimensional characteristics. Essentially, the significant distortion arises from interlayer ordering, adding complexity to the Klein tunneling process~\cite{44}. A consensus on a 2$\times$2 reconstruction in the kagome plane is established; however, comprehensive calculations are essential to determine the exact periodic units along the \textit{c} axis.\\

\section{\label{sec:level4}Conclusion}
In summary, we conducted comprehensive theoretical and experimental analyses of the FS structure in the layered kagome material CsV$_{3}$Sb$_{5}$. Our study reveals several distinct frequencies ranging from 19\,T to 2021\,T, which agree well with theoretical predictions for FSs. The observed the combinational and forbidden frequencies in the Fourier spectrum unambiguously highlight the features associated with magnetic breakdown. The limited el-ph effects are in agreement with weak correlations. Angular-dependence dHvA oscillation data for ground state electron behavior demonstrate quantitative consistency with multifold 2$\times$2$\times$2 distortion mode in CsV$_{3}$Sb$_{5}$. Finally, the selection of calculation distortion is worth further exploring, which could deepen our understanding of this complex layered kagome material.\\

\begin{acknowledgments}
We acknowledge very helpful discussions with Prof.~ZiJi Xiang. This work was financially supported by the National Key R$\&$D Program of the MOST of China (Grant No. 2022YFA1602602, 2022YFA1602603), the National Natural Science Foundation of China (Grants No. 12122411, 12104459), The Basic Research Program of the Chinese Academy of Sciences Based on Major Scientific Infrastructures (Grants No.JZHKYPT-2021-08), Excellent Program of Hefei Science Center CAS (Grant No. 2021HSC-CIP016), The $\text{ CASHIPS Director' Fund}$ (Grant No. YZJJ2022QN36). 
\end{acknowledgments}


\renewcommand{\newblock}{\References}

\end{document}